\documentstyle[twoside,psfig]{article}

\catcode`\@=11
\long\def\@makefntext#1{
\protect\noindent \hbox to 3.2pt {\hskip-.9pt
$^{{\eightrm\@thefnmark}}$\hfil}#1\hfill}               %CAN BE USED

\def\@makefnmark{\hbox to 0pt{$^{\@thefnmark}$\hss}}    %ORIGINAL

\def\ps@myheadings{\let\@mkboth\@gobbletwo
\def\@oddhead{\hbox{}
\rightmark\hfil\eightrm\thepage}
\def\@oddfoot{}\def\@evenhead{\eightrm\thepage\hfil
\leftmark\hbox{}}\def\@evenfoot{}
\def\sectionmark##1{}\def\subsectionmark##1{}}

\oddsidemargin=\evensidemargin
\newcounter{sectionc}\newcounter{subsectionc}\newcounter{subsubsectionc}
\renewcommand{\section}[1] {\vspace{12pt}\addtocounter{sectionc}{1}
\setcounter{subsectionc}{0}\setcounter{subsubsectionc}{0}\noindent
        {\tenbf\thesectionc. #1}\par\vspace{5pt}}
\renewcommand{\subsection}[1] {\vspace{12pt}\addtocounter{subsectionc}{1}
      \setcounter{subsubsectionc}{0}\noindent
      {\bf\thesectionc.\thesubsectionc.{\kern1pt \bfit #1}}\par\vspace{5pt}}
\renewcommand{\subsubsection}[1]
      {\vspace{12pt}\addtocounter{subsubsectionc}{1}
      \noindent{\tenrm\thesectionc.\thesubsectionc.\thesubsubsectionc.
      {\kern1pt \tenit #1}}\par\vspace{5pt}}
\newcommand{\nonumsection}[1] {\vspace{12pt}\noindent{\tenbf #1}
        \par\vspace{5pt}}

\newcounter{appendixc}
\newcounter{subappendixc}[appendixc]
\newcounter{subsubappendixc}[subappendixc]
\renewcommand{\thesubappendixc}{\Alph{appendixc}.\arabic{subappendixc}}
\renewcommand{\thesubsubappendixc}
        {\Alph{appendixc}.\arabic{subappendixc}.\arabic{subsubappendixc}}

\renewcommand{\appendix}[1] {\vspace{12pt}
        \refstepcounter{appendixc}
        \setcounter{figure}{0}
        \setcounter{table}{0}
        \setcounter{lemma}{0}
        \setcounter{theorem}{0}
        \setcounter{corollary}{0}
        \setcounter{definition}{0}
        \setcounter{equation}{0}
        \renewcommand{\thefigure}{\Alph{appendixc}.\arabic{figure}}
        \renewcommand{\thetable}{\Alph{appendixc}.\arabic{table}}
        \renewcommand{\theappendixc}{\Alph{appendixc}}
        \renewcommand{\thelemma}{\Alph{appendixc}.\arabic{lemma}}
        \renewcommand{\thetheorem}{\Alph{appendixc}.\arabic{theorem}}
        \renewcommand{\thedefinition}{\Alph{appendixc}.\arabic{definition}}
        \renewcommand{\thecorollary}{\Alph{appendixc}.\arabic{corollary}}
        \renewcommand{\theequation}{\Alph{appendixc}.\arabic{equation}}
        \noindent{\tenbf Appendix \theappendixc #1}\par\vspace{5pt}}
\newcommand{\subappendix}[1] {\vspace{12pt}
        \refstepcounter{subappendixc}
        \noindent{\bf Appendix \thesubappendixc. {\kern1pt \bfit #1}}
        \par\vspace{5pt}}
\newcommand{\subsubappendix}[1] {\vspace{12pt}
        \refstepcounter{subsubappendixc}
        \noindent{\rm Appendix \thesubsubappendixc. {\kern1pt \tenit #1}}
        \par\vspace{5pt}}

\newcommand{\smalllineskip}{\baselineskip=10pt}

\def\eightcirc{
\begin{picture}(0,0)
\put(4.4,1.8){\circle{6.5}}
\end{picture}}
\def\eightcopyright{\eightcirc\kern2.7pt\hbox{\eightrm c}}

\def\abstracts#1#2#3{{
        \centering{\begin{minipage}{4.5in}\baselineskip=10pt\footnotesize
        \parindent=0pt #1\par
        \parindent=15pt #2\par
        \parindent=15pt #3
        \end{minipage}}\par}}

\renewenvironment{thebibliography}[1]
        {\frenchspacing
         \ninerm\baselineskip=11pt
         \begin{list}{\arabic{enumi}.}
        {\usecounter{enumi}\setlength{\parsep}{0pt}
         \setlength{\leftmargin 12.7pt}{\rightmargin 0pt} %FOR 1--9 ITEMS
         \setlength{\itemsep}{0pt} \settowidth
        {\labelwidth}{#1.}\sloppy}}{\end{list}}

\newcounter{itemlistc}
\newcounter{romanlistc}
\newcounter{alphlistc}
\newcounter{arabiclistc}

\newcommand{\fcaption}[1]{
        \refstepcounter{figure}
        \setbox\@tempboxa = \hbox{\footnotesize Fig.~\thefigure. #1}
        \ifdim \wd\@tempboxa > 5in
           {\begin{center}
        \parbox{5in}{\footnotesize\smalllineskip Fig.~\thefigure. #1}
            \end{center}}
        \else
             {\begin{center}
             {\footnotesize Fig.~\thefigure. #1}
              \end{center}}
        \fi}

\newcommand{\tcaption}[1]{
        \refstepcounter{table}
        \setbox\@tempboxa = \hbox{\footnotesize Table~\thetable. #1}
        \ifdim \wd\@tempboxa > 5in
           {\begin{center}
        \parbox{5in}{\footnotesize\smalllineskip Table~\thetable. #1}
            \end{center}}
        \else
             {\begin{center}
             {\footnotesize Table~\thetable. #1}
              \end{center}}
        \fi}

\def\@citex[#1]#2{\if@filesw\immediate\write\@auxout
        {\string\citation{#2}}\fi
\def\@citea{}\@cite{\@for\@citeb:=#2\do
        {\@citea\def\@citea{,}\@ifundefined
        {b@\@citeb}{{\bf ?}\@warning
        {Citation `\@citeb' on page \thepage \space undefined}}
        {\csname b@\@citeb\endcsname}}}{#1}}

\newif\if@cghi
\def\cite{\@cghitrue\@ifnextchar [{\@tempswatrue
        \@citex}{\@tempswafalse\@citex[]}}
\def\citelow{\@cghifalse\@ifnextchar [{\@tempswatrue
        \@citex}{\@tempswafalse\@citex[]}}
\def\@cite#1#2{{$\null^{#1}$\if@tempswa\typeout
        {IJCGA warning: optional citation argument
        ignored: `#2'} \fi}}

\def\@refcitex[#1]#2{\if@filesw\immediate\write\@auxout
        {\string\citation{#2}}\fi
\def\@citea{}\@refcite{\@for\@citeb:=#2\do
        {\@citea\def\@citea{, }\@ifundefined
        {b@\@citeb}{{\bf ?}\@warning
        {Citation `\@citeb' on page \thepage \space undefined}}
        \hbox{\csname b@\@citeb\endcsname}}}{#1}}

\def\@refcite#1#2{{#1\if@tempswa\typeout
        {IJCGA warning: optional citation argument
        ignored: `#2'} \fi}}

\def\refcite{\@ifnextchar[{\@tempswatrue
        \@refcitex}{\@tempswafalse\@refcitex[]}}

\def\pmb#1{\setbox0=\hbox{#1}
        \kern-.025em\copy0\kern-\wd0
        \kern.05em\copy0\kern-\wd0
        \kern-.025em\raise.0433em\box0}

\def\fnt#1#2{\footnotetext{\kern-.3em
        {$^{\mbox{\scriptsize #1}}$}{#2}}}

\def\fpage#1{\begingroup
\voffset=.3in
\thispagestyle{empty}\begin{table}[b]\centerline{\footnotesize #1}
        \end{table}\endgroup}

\def\runninghead#1#2{\pagestyle{myheadings}
\markboth{{\protect\footnotesize\it{\quad #1}}\hfill}
{\hfill{\protect\footnotesize\it{#2\quad}}}}
\font\tenrm=cmr10
\font\tenit=cmti10
\font\tenbf=cmbx10
\font\bfit=cmbxti10 at 10pt
\font\ninerm=cmr9

\font\eightrm=cmr8

\def\qed{\hbox{${\vcenter{\vbox{                      %HOLLOW SQUARE
   \hrule height 0.4pt\hbox{\vrule width 0.4pt height 6pt
   \kern5pt\vrule width 0.4pt}\hrule height 0.4pt}}}$}}

\begin{document}

\runninghead{Yu.A. Baurov}
{Structure of physical space$\ldots$}

%\normalsize\textlineskip
\thispagestyle{empty}\setcounter{page}{1}
\vspace*{0.88truein}
\fpage{1}

\centerline{\bf STRUCTURE OF PHYSICAL SPACE}
\centerline{\bf AND NEW INTERACTION IN NATURE}
\centerline{\bf (THEORY AND EXPERIMENTS)}

\vspace*{0.035truein}

\vspace*{0.37truein}
\centerline{\footnotesize Yuri A. Baurov}

\centerline{\footnotesize \it
Central Research Institute of Machine Building}
\baselineskip=10pt
\centerline{\footnotesize \it
Pionerskaya, 4}
\baselineskip=10pt
\centerline{\footnotesize \it
141070, Korolyov, Moscow Region, Russia}

\baselineskip 5mm

\vspace*{0.21truein}

\abstracts{In the talk, on the basis of the author's model of formation of the observable physical space $R_3$
in the process of dynamics of special discrete one-dimensional vectorial objects, {\it byuons}, while minimizing their
potential energy of interaction in the one-dimensional space $R_1$ formed by them, the existence of global
anisotropy of observable space and new interaction in nature, is shown. The data of recent experiments are given and discussed.}{}{}
\bigskip
\section{\bf Introduction.}\vskip10pt
A wealth of works, beginning from antiquity (Aristoteles, Euclides, Democritus) and ending with
authors of $X\!X$-th century, are dedicated to the structure of space and time, to the physical sense of these
fundamental concepts, and their properties\cite{1}. In all existing works on quantum field theory and
physics of elementary particles, the space in which elementary processes occur, as a rule, is given one
way or another. Yet we will follow another way and try to build physical space and major properties of
elementary objects in this space from dynamics of a finite set of special discrete objects (so called
{\it byuons})\cite{2}. Note that the development of physical comprehension of elementary processes on the base
of modern superstring models\cite{3}, unfortunately, also gives no evidences for how structured is the
observed space itself which is obtained, according to one of the models, by means of compactification
of six dimensions in a ten-dimensional space. New theoretical approaches found in construction of
physical space have given the chance to look in a new way also at the most studied object of the
classic and quantum field theory, the electromagnetic field\cite{4}.
In the present paper we consider only basic statements of the byuon theory as well as the new
interaction connected with the existence of the cosmological vectorial potential, a new fundamental
vectorial constant entering into the definition of byuons. From a wealth of experimental material on
investigating the new interaction\cite{5}${}^-$\cite{67}, we shall briefly consider only the influence of the new
interaction on the rate of the $\beta$-decay of radioactive elements\cite{67}.

\section{\bf Fundamental Theoretical Concepts of Physical Space Origin
and of New Force.}\vskip10pt
In Ref.\cite{2}, the conception of formation of the observed
physical space $R_3$ from a finite set of byuons is given. The byuons
${\bf\Phi}(i)$ are one-dimensional vectors and have the form:
$${\bf\Phi}(i) ={\bf A_g}x(i),$$
where $x(i)$ is the "length" of the byuon, a real (positive, or negative)
value depending on the index $i=0,1,2,...,k...$, a quantum number of
${\bf\Phi}(i)$; under $x(i)$ a certain time charge of the byuon may be
meant (with $x(i)$ in centimeters).  The vector ${\bf A}_g$  represents the
cosmological vectorial potential, a new basic vectorial constant
\cite{2}. It takes only two values:
$$ {\bf A_g} = \left\{ A_g\atop-\sqrt{-1} A_g\right\} ,$$
where $A_g$ is the modulus of the cosmological vectorial potential
($ A_g \approx 1.95\times 10^{11}$ CGSE units).  According to the theory
\cite{2}, the value $A_g$ is the limiting one. In reality, there exists in
nature, in the vicinity of the Earth, a certain summary
potential, since the vectorial potential fields from the Sun
($A_\odot \approx 10^8$ CGSE units), the Galaxy ($ \sim 10^{11}$ CGSE
units), and the Metagalaxy ($ > 10^{11}$ CGSE units) are superimposed on the constant $A_g$
resulting probably in some turning of ${\bf A}_\Sigma$  relative to the
vector ${\bf A_g}$ in the space $R_3$ and in a decrease of it.

Hence in the theory of physical space (vacuum) which the present article leans upon,
the field of the vectorial potential introduced even by Maxwell
gains a fundamental character. As is known, this field was
believed as an abstraction. All the existing theories are
usually gauge invariant. For example, in classical and
quantum electrodynamics, the vectorial potential is defined with
an accuracy of an arbitrary function gradient, and the scalar
potential is with that of time derivative of the same function, and
one takes only the fields of derivatives of these potentials, i.e.
magnetic flux density and electric field strength, as real.

ln Refs.\cite{7}${}^-$\cite{11},%[11-15]
 local violation of the gauge invariance and
Poincaret's group was supposed, and the elementary particle
charge and quantum number formation processes were investigated
in some set, therefore the potentials gained an unambiguous
physical meaning there. In the present paper, this is a finite
set of byuons. The works by D.Bohm and Ya.Aharonov %\cite{18}
discussing the special meaning of potentials in quantum
mechanics are the most close to the approach under consideration.
%they are confirmed by numerous experiments \cite{19}.

The byuons may be in four  vacuum states (VS)
$II^+, I^+, II^-, I^-,$ in which they discretely change
the value ot $x(i)$: the state $II^+$ discretely increases ($c = c_0 =
\tilde x_0/\tau_0$,
where $\tilde x_0$ - quantum of space ($\approx 10^{-33}$cm), $\tau_0$ -
quantum
of time ($\approx 10^{-43}c$)) and $I^+$ decreases $x(i)>0$ ($c = -c_o =
-\tilde x_0/\tau_0$);
the states $II^-$ and $I^-$ discretely increase
or decrease the modulus of $x(i)<0$, respectively ($II^-$ corresponds to $c
= - c_0$,
$I^-$ corresponds to $c = c_0$). The sequence of discrete changes of $x(i)$
value is defined as a proper discrete time of the byuon.
The byuon vacuum states originate randomly\cite{2}.

In Ref.\cite{2}, the following hypothesis has been put forward:
\vskip10pt
{\it It is suggested that the observed space $R_3$ is built up as a result of
minimizing the potential energy of byuon interactions in the one-dimensional space
$R_1$ formed by them. More precisely, the space $R_3$ is
fixed by us as a result of dynamics arisen of byuons. The dynamic processes and, as a consequence,
wave properties of elementary particles appear therewith in the space $R_3$ for objects with residual
positive potential energy of byuon interaction (objects observed).}
\vskip10pt
Let us briefly list the results obtained earlier  when investigating the present model of physical
vacuum.

1. The existence of a new long-range
interaction in nature, arising when acting on physical vacuum by
the vectorial potential of high-current systems, has been
predicted\cite{5}.

2. All the existing interactions (strong,
weak, electromagnetic and gravitational ones) along with the new
interaction predicted have been qualitatively explained in the
unified context of changing in three periods of byuon
interactions $k,N,P$ with characteristic scales - $x_0 = \tilde x_0 k \approx
10^{-17}$cm, $ct^* = \tilde x_0 kN \approx 10^{-13}$cm, and $\tilde x_0 kNP \approx
10^{28}$cm, determined from the minimum potential energy of byuon interaction\cite{2}.

3. Masses of leptons, basic barions and mesons have been found\cite{10,11}.

4. The constants of weak interaction (vectorial and axial ones)
and of strong interaction have been calculated\cite{10,11}.

5. The origin of the galactic and intergalactic magnetic fields has
been explained as a result of existence of an insignificant
($\approx 10^{-15}$) asymmetry in the formation process of $R_3$ from the
one-dimensional space of byuons\cite{2}.

6. The matter density observed in the Universe ($\approx 10^{-29} g/cm^3$)
has been computated\cite{2}.

7. The origin of the relic radiation has been explained on the
basis of unified mechanism of the space $R_3$ formation from
one-dimensional space $R_1$ of byuons\cite{2}.

8. ${\bf A_g}$-vector has coordinates $\alpha\approx 270^\circ, \delta\approx 34^\circ$
in the second equatorial system\cite{5}${}^-$\cite{67}.

Let us explain item 1 briefly. It is shown in Ref. \cite{2,11} that masses of
all elementary particles are proportional to the modulus of ${\bf A_g}$.
If we direct now the vectorial potential of a magnetic system in
some space region towards the vector ${\bf A_g}$ then any material body
will be forced out of the region where $|{\bf A}_\Sigma | < |{\bf A_g}|$ .
The new force is nonlocal and nonlinear.
This force is directed mainly by the direction of the
vector ${\bf A_g}$, but as the latest experiments have shown, there is also an
isotropic component of the new force.

It can be shown that the magnituge
$$F\sim\Delta A\cdot\frac{\partial\Delta A}{\partial X}.$$
We will take $\Delta A$ equal to the difference in
$|{\bf A}_\Sigma|$ changes due to the source of magnetic field at the location points of a weight (test body)
%$|{\bf A}_\Sigma|_w$
and sensing element;
%$|{\bf A}_\Sigma|_d$;
$\frac{\partial\Delta A}{\partial X}$ is gradient of $\Delta A$ at $R_3$.

One of the important predictions of the theory is revealing a new information channel in the
Universe which is associated with the existence of a minimum object with positive potential energy, so
called object $4b$, arising in the minimum four-contact interaction of byuons in the vacuum states
$II^+, I^+, II^-, I^-$.
Object $4b$ may be identified with the pair of the neutrino-antineutrino
($\nu_e \Leftrightarrow \tilde \nu_e$) In four-contact byuon
interaction, a minimum action equal to $h$ (Planck's constant)
occurs, and the spin of the object appears. Hence the greater
part of the potential energy of byuon interaction is transformed
into the spin of the object $4b$. The residual (after minimization)
potential energy of the object $4b$ is equal to $\approx 33eV$. It is
identified with the rest mass of this object in the space $R_1$.
In agreement with Refs.\cite{2}, the indicated minimum object $4b$
has, according to Heisenberg uncertainty relation, the
uncertainty in coordinate $\Delta x\approx 10^{28}$cm in $R_3$.
The total energy of these objects determines near $98\%$ energy of the
Universe as well as its matter density observed (dark matter).

%In Refs. \cite{4,5} it is shown
%that the existence of the $4b$-objects interpreted by the author
%as pairs of electron-type neutrino and electron-type
%antineutrino ($\nu_e \Leftrightarrow \tilde \nu_e$), causes a finiteness
%of the velocity of propagation of interaction c being an infinite
%quantity at $t = 0$  and  $t = t^*$ according to Eq. (11).

\section{\bf Experimental Investigatios on Influence of Vectorial Potential of
Magnetic Systems on $\beta$-decay Count Rate.}\vskip10pt
It was shown\cite{2,11} that the essence of the weak interactions consists in extending the process
of forming the electrical charge of the particle over minimum distance of $\sim 2x_0$, or, in terms of the
theory of $R_3$ space formation from byuons, over two periods of byuon interaction, i.e. over
$2x_0 = 2k\tilde x_0$. The "extension" is in the direction of the vector ${\bf A}_g$. For the model of weak four-fermion
interaction with a current taken in the form of the sum of weak vectorial current $V'$ and weak axial
(psevdovectorial) current $A'$ ($V'-A'$ - interaction), the interaction constants $C_V$ and $C_A$ have been
determined in terms of the new ideology\cite{2,11} as $|C_{V,A}|\sim 1/|{\bf A}_g|^2$. Since the
energy of decay electrons $E_\beta \sim |A_\Sigma|$, and the probability of $\beta$-decay\cite{12}
$W \sim |C_{V,A}|^2 E_\beta^5$ in the framework of $V'-A'$ - interaction, we have
$W \sim |A_\Sigma|$.

Thus, as $|A_\Sigma|$ changes under the influence of vectorial potential of a magnetic system, one might
expect $W$ being also changed.

In the existing pulsed solenoidal systems one can achieve values of the vectorial magnetic
potential $A_g$ at a level of about $10^6 - 10^7 Gs\cdot cm$, i.e. the deviation of the decay number $N$ from the initial
value of registered decay events $N_0$ will not exceed $N \sim A N_0 / A_g = 10^{-5}N_0$.
As is shown above, the vectorial magnetic potentials of the Earth's and Sun's magnetic dipoles can exceed the above
mentioned value of $A$ by one-two orders. In this case one may expect deviations of the decay number
at a level of $10^{-3}$ from the undisturbed value, what substantially simplifies the conditions of
measurements for detecting the effect.

As the expression for the new force contains not only $\Delta A$ but $\frac{\partial\Delta A}{\partial X}$,
too, one may assume the variation of probability $\Delta W$ of $\beta$-decay to be proportional to $\Delta A\cdot\frac{\partial\Delta A}{\partial X}$.
As is known\cite{12}, the neutron has a magnetic moment $M_n \approx 10^{-23} erg/Gs$. Scale estimations show
that in connection with the presence of $M_n$, we can say about the values of $\frac{\partial\Delta A}{\partial\ X} \approx 10^{16}Gs\cdot cm$
and $\Delta A \approx 10^3 Gs\cdot cm$ in the vicinity of a neutron. Because of the rotation of the Earth and the action of the
Sun's magnetic system, $\Delta A$ can vary in time over five orders of magnitude and more. Hence we may
assume, that the variation of $W$ due to the action of the new force can be observable.The experiment
was principally based on a search for 24-hour's periods of oscillations in intensity of $\beta$-decay caused
by the action of the vectorial potential ${\bf A_E}$, and on the fixation of these periods in spacial coordinates.
By now six runs of experiments have been performed to verify this assumption. We shall consider
only one run of experiments (August-September, 1996)\cite{67}.

\nonumsection{\bf Experiment}\vskip10pt
The experiment was carried out in the Laboratory for nuclear reactions (FLNR) named after
G.N.Flerov, at JINR (Dubna)\cite{67}. The measurements were made by the scintillation procedure
with a ${}^{90}Sr$-source.

\nonumsection{\bf Analysis of Experiments}\vskip10pt
The results are shown in Figs.1-3. The standard Fourier analysis of the series (Fig.1) was carried
out. In Fig.2 where the results are presented, at least two frequencies stand out, of which the former
("great peak") corresponds to approximately half-week period, and the latter ("small peak") does to
the daily one. The daily periodicity found by Fourier-method gives, however, no way to determine the
astronomical time of an event when the measured value is greater or less than the average.
The following procedure was used to refine the distribution of $\beta$-decay numbers over the astronomical
day. Each moment (minute) of measurement was represented as a point of a circle (corresponding to
an astronomical day) and expressed in degrees, so that the whole series could be "coiled" around that
circle. Thus, each measurement was related to a certain time of day (in degrees). If the quantity to be
measured is isotropic in time, then the distribution thereof over the circle will be uniform, and the
hypothesis for uniformity of said distribution may be verified by statistical methods. For this purpose
we used the Kolmogoroff-Smirnoff's test.

According to our conception of influence of changes in $A_\Sigma$ on the $\beta$-decay  count rate, we are
interested primarily in values of fluxes lesser than the average indication of the instrument. On this
basis, when analyzing a distribution, we take into account only the values $L\times S$ lesser than the average
where $S$ is the standard deviation of the entire series (Fig.1), and $L$ is a factor determining the extent
of deviation from the average. For each of such an extreme value, one notes the point in time at which
that takes place, and tests the hypothesis for uniformity of distribution of those points over the
astronomical day expressed in degrees (24 hours are equal to $360^\circ$). In Fig.3
the results of computation for $L = 2$ are graphically represented. The $X$ axis is astronomical time in degrees
(from $0^\circ$ to $360^\circ$), the $Y$ axis is deviation from the uniform distribution.
The confidence interval for $P=0.05$ is
shown dashed. It is clearly seen that the frequency function of the sample is highly nonuniform and
peaks at an angle of about $90^\circ$. At this point, the $5\%$ level of significance is far exceeded.

Since the counting in this experiment began on 23rd August, 1996, at 20 hs according to Moscow
time (i.e. at 18 hs of astronomical time), the angle of $90^\circ$ corresponds to 24 hs of astronomical
time. In Fig.4 the indicated point of time is asterisked. At the starred point (*), a tangent to the line of
vectorial potential ${\bf A_E}$ of the Earth's magnetic field is drawn (Fig.4). The tangent forms an angle of
$\sim 30^\circ$ with the assumed direction of ${\bf A_g}$. In the Fig.4 the similar asterisks and tangent lines indicate
points of extremum deviations in variations of $W$ for the run of experiments carried out before 1999.
The astronomical times of extremums are shown, too. As is seen from the Fig.4, the tangents to the
line of the vectorial potential ${\bf A_E}$ discern two directions making an angle around $30^\circ-45^\circ$ with the line
of ${\bf A_g}$ to the left or right of it. The obtained result corresponds to our theoretical views of the action of
the new force.

The experiment in JINR goes on uninterruptedly since 09.12.1998. The results of February and
March, 1999, are coincident with those of 1998 with an accuracy of $5-10^\circ$. It is interesting also to note
that the results of experiments carried out in April, 1998, are coincident with the experiments
performed in April, 1994 by another collective of authors\cite{67}. Before long, they will be presented to
the journal "Modern Physics" as a collective work.

\nonumsection{\bf ACKNOWLEDGEMENTS.}

The author is grateful to S.T.Belyaev, an academician of RAS, Yu. Ts. Oganesian, a
corresponding member of RAS for the support of the work and fruitful discussions, to Yu.G.Sobolev,
V.F.Kushniruk for the performance of the experiments, and to A.A.Konradov for the processing of the
results.

\nonumsection{References}

\begin{figure}[thb]  %fig.1
\centerline{\psfig{figure=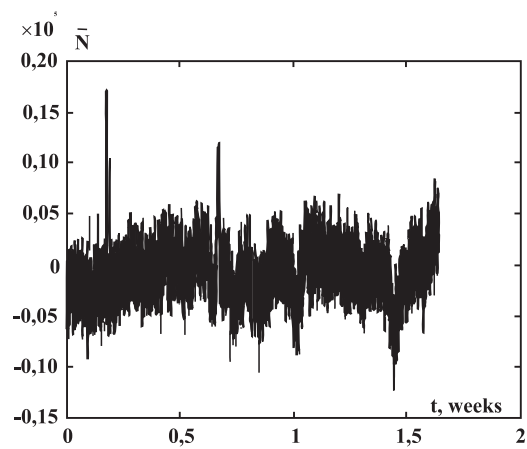}}%,angle=0,height=60mm}}
\vspace{5mm}
\caption{The  normalized  count rate averaged  over  1  minute
versus  time after extracting of the nonlinear  trend  from
the initial series of measurements.}
\end{figure}
%%%%%%%%%%%%%%%%%%%%%%%%%%%%%%%%%%%%%%%%%%%%%%%%%%%%%%%%%%%%%%%%%%%
\begin{figure}[hb]  %fig.2
\centerline{\psfig{figure=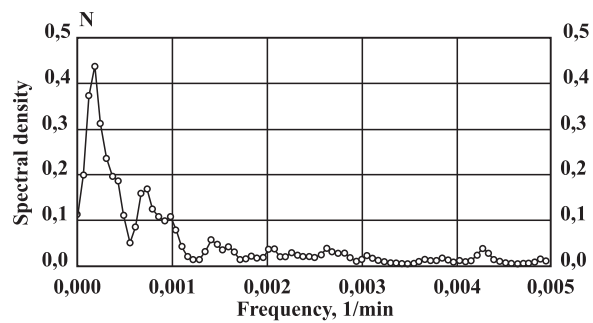}}%,angle=0,height=60mm}}
\vspace{5mm}
\caption{The Fourier spectrum of the signal shown in Fig.3.
The  first  peak  corresponds  to   half-week
period, the second one corresponds to 24-hour period.}
\end{figure}
%%%%%%%%%%%%%%%%%%%%%%%%%%%%%%%%%%%%%%%%%%%%%%%%%%%%%%%%%%%%%%%%%%%
\begin{figure}[thb]  %fig.3
\centerline{\psfig{figure=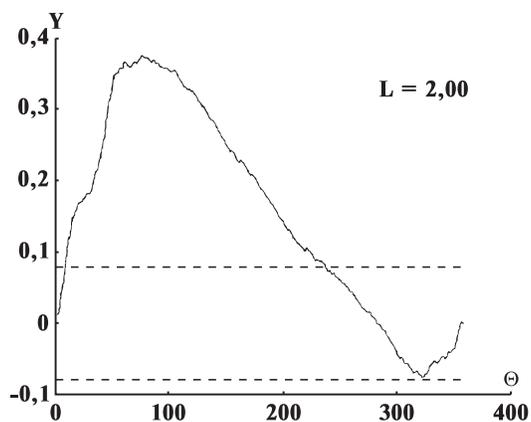}}%,angle=0,height=60mm}}
\vspace{5mm}
\caption{
The dynamics of the ununiformity for
a  24-hour period. $Y$  is the Kolmogorov statistic value.
}
\end{figure}
%%%%%%%%%%%%%%%%%%%%%%%%%%%%%%%%%%%%%%%%%%%%%%%%%%%%%%%%%%%%%%%%%%%%
\begin{figure}[hb]  %fig.4
\centerline{\psfig{figure=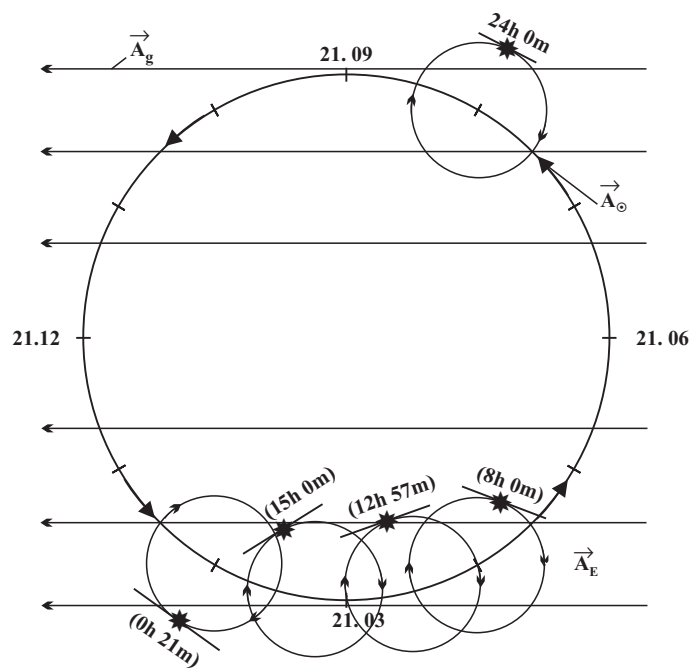}}
%,angle=0,height=60mm}}
\vspace{5mm}
\caption{The   directions  of  the  ${\bf A}_g$  vector  and   vectorial
     potentials   of   magnetic  fields   from   the   Sun  ${\bf A}_\odot$
     and the Earth ${\bf A}_E$. $8^h$, (*) - astronomical time points corresponding to
extremum changes in $\beta$-decay count rate.}
\end{figure}

\end{document}